\documentclass[english]{article}
\usepackage[T1]{fontenc}
\usepackage[utf8]{inputenc}
\usepackage{amsmath}
\usepackage{amssymb}
\usepackage{esint}
\usepackage{babel}
\usepackage[auth-lg]{authblk}
\begin{document}

\title{Generalized calculus in radiobiology: Physical implications}
\author[1]{O. Sotolongo-Grau}
\author[1]{D. Rodriguez-Perez}
\author[1,2]{J. C. Antoranz}
\author[2]{Oscar Sotolongo-Costa}
\affil[1]{\small{Departamento de Física Matemática y de Fluidos, Universidad Nacional de Educación a Distancia (UNED), Spain}}
\affil[2]{\small{Cátedra de Sistemas Complejos “Henri Poincaré”, Universidad de La Habana, Cuba}}
\date{~}
\maketitle
\begin{abstract}
Non-extensive statistical physics has allowed to generalize mathematical
functions such as exponential and logarithms. The same framework is
used to generalize sum and product so that the operations allow a
more fluid way to work with mathematical expressions emerging from
non-additive formulation of statistical physics. In this work we employ
the generalization of the exponential, logarithm and product to obtain
a formula for the survival fraction corresponding to the application
of several radiation doses on a living tissue. Also we provide experimental
recommendations to determine the universal characteristics of living
tissues in interaction with radiation. These results have a potential
application in radiobiology and radiation oncology.
\end{abstract}
Entropy, Statistical mechanics, Radiation, Non-extensivity
\section{Introduction}

Biological systems are the paradigm of open non equilibrium systems
characterized by long range correlations and by constantly occurring
non markovian processes. This makes its description in terms of equilibrium
statistical physics a very hard task. Thus a formulation of processes
in this kind of systems must be made in a non orthodox statistical
framework in which the concept of entropy is expressed in a more general
way. 

Tsallis's definition of entropy has been successfully used, in the
last years, to extend the classical statistical mechanics to several
problems involving very complex correlations and properties. Its expression
for the entropy (in units of the Boltzmann's constant) \cite{Tsallis98}
is\begin{equation}
S_{q}=\frac{1}{q-1}\left(1-\int_{0}^{\infty}p^{q}(x)dx\right),\label{eq:tsallis_ent}\end{equation}
where $p(x)dx$ is the probability of the magnitude $x$ to adopt
values between $x$ and $x+dx$ and $q$ is the {}``index of non-extensivity''.
This, together with the constraints that the probability density satisfies
the normalization condition and the finiteness of the {}``$q$-mean
value'' $\int_{0}^{\infty}p^{q}(x)xdx=\left\langle x\right\rangle _{q}<\infty$,
has been applied in different problems with excellent results even
when statistical physics based on Boltzmann-Gibbs (BG) entropy had
not succeed. On the other hand, Tsallis entropy reduces to the BG
one when $q=1$, so that in this sense the former entropy can be interpreted
as a generalization of the last one.

Our objective in this work is to employ this new formulation of statistical
physics and the consequent generalization of several mathematical
functions and operations, provided in the framework of this new formulation,
to obtain a simple universal formula for the survival fraction of
tissue cells under radiation.

We apply the maximum entropy principle using the Tsallis entropy to
obtain an expression for the survival fraction in terms of the radiation
dose. Then using a generalization of the definition of logarithm and
exponential function and a generalized expression for the product
the resultant survival fraction of several radiation sessions is obtained.

\section{Materials and methods}

To apply the maximum entropy principle in the Tsallis version to the
problem of finding the survival factor of a living tissue \cite{STEEL}
that receives a radiation, we postulate the existence of some amount
of absorbed radiation $\Delta<\infty$(or its equivalent {}``limit
effect'', $\Omega=\alpha_{0}\Delta$) after which no cell survives.
The application of the maximum entropy principle performs like the
usual one but with a few modifications.

The Tsallis entropy becomes:\begin{equation}
S_{q}=\frac{1}{q-1}\left(1-\int_{0}^{\Omega}p^{q}(E)dE\right),\label{eq:Tsallis_rent}\end{equation}
where $p(E)$ is the cell death probability, dependent on the {}``radiation
effect'' $E$, which is a function of the applied dose $D$. The
way $E$ depends on $D$ is a controversial matter, being generally
accepted a linear relation $E=\alpha D$ and also the {}``linear
quadratic model'' $E=\alpha D+\beta D^{2}$.

The normalization condition is in this case $\int_{0}^{\Omega}p(E)dE=1$
and the $q$-mean value becomes $\int_{0}^{\Omega}p^{q}(E)EdE=\left\langle E\right\rangle _{q}<\infty$.
With this definition, all properties of the tissue and its characteristics
of the interaction with radiation become included in $\left\langle E\right\rangle _{q}$
and therefore in $\Omega$. This is the only parameter (besides $q$)
entering in our description. It is clear that the determination of
$\left\langle E\right\rangle _{q}$ for the different tissues under
different conditions of radiation would give the necessary information
for the characterization of the survival factor.

To calculate the maximum of \eqref{eq:Tsallis_rent} under the above
conditions the well known method of Lagrange multipliers \cite{Plastino99}
is applied, obtaining\begin{equation}
\Omega=\frac{2-q}{1-q}\left(\frac{\left\langle E\right\rangle _{q}}{2-q}\right)^{\frac{1}{2-q}}\label{eq:omega_a}\end{equation}
and\begin{equation}
p(E)=\left(\frac{2-q}{\left\langle E\right\rangle _{q}}\right)^{\frac{1}{2-q}}\left(1-\frac{1-q}{2-q}\left(\frac{2-q}{\left\langle E\right\rangle _{q}}\right)^{\frac{1}{2-q}}E\right)^{\frac{1}{1-q}}.\label{eq:lmT}\end{equation}

Then the survival factor is\begin{equation}
F_{s}(E)=\int_{E}^{\Omega}p(x)dx=\left(1-\frac{E}{\Omega}\right)^{\frac{2-q}{1-q}}\label{eq:psD}\end{equation}
with $q<1$ for $E<\Omega$ and zero otherwise. It is not hard to
see that when $q\rightarrow1$ then $\Omega\rightarrow\infty$ and
$\left\langle E\right\rangle _{q}\rightarrow\left\langle E\right\rangle $.

Equation \eqref{eq:psD} can be written \begin{equation}
F_{s}(D)=\begin{cases}
\left(1-\frac{D}{\Delta}\right)^{\gamma} & \forall D<\Delta\\
0 & \forall D\geqslant\Delta\end{cases}\label{eq:finalF}\end{equation}
where we introduced $E=\alpha_{0}D$ and $\gamma=\frac{2-q}{1-q}$.

Equation \eqref{eq:finalF} represents the survival fraction in terms
of the measurable quantities $D$ (radiation dose) and $\Delta$.
All the information about the kind of radiation, radiation rate, etc
is contained in the phenomenological term $\Delta$, whereas tissues
are characterized by $\gamma$. This makes \eqref{eq:finalF} a very
general expression with universal characteristics since the phase
transition described by \eqref{eq:finalF} is homomorphic with the
phase transition of ferromagnets near the Curie point. The exponent
$\gamma$ in this case, as in ferromagnetic phase transitions, determines
the universality class. Then $\gamma$ in our case deals only with
the kind of tissue that interacts with radiation \cite{sotolongograu-2009}.

The value of $\Delta$ characterizes a critical point for cell survival
probability such that for $D<\Delta$ the probabilities of cell survival
and death coexist but when $D>\Delta$ a phase transition takes place
and no cell survives. Note that we have postulated a linear relation
between $E$ and $D$. With this, the curious property that the tissue
effect $E$ is additive while the survival fraction is not multiplicative
emerges.

The linear model for the tissue effect \cite{Tubiana}, however, gives
that if the dose is additive the corresponding survival fraction is
multiplicative if their mean values are the same. Though this property
belongs only to the linear model and not to more general descriptions
like the LQ model \cite{Tubiana} and others, we think it is worth
to find a link between the additivity property of the dose and the
probabilistic properties of the cell survival fraction.

To motivate this let us first observe that the Tsallis definition
of entropy, though gives a more general expression than the BG one,
lacks one of the important properties of this entropy: extensivity.

Indeed, let us consider a system composed by two independent subsystems
$A$ and $B$. Extensivity of the entropy means that the entropy of
the whole system is the sum of the entropies of the subsystems: $S(A\cup B)=S(A)+S(B)$.
But in the Tsallis case $S_{q}(A\cup B)=S_{q}(A)+S_{q}(B)+(q-1)S_{q}(A)S_{q}(B)$. 

As Tsallis entropy can be defined through the introduction of the
$q$-logarithm: \begin{equation}
S_{q}=\int_{0}^{\infty}p^{q}(x)lq\left(p(x)\right)dx\label{eq:Tent}\end{equation}
where $lq(x)=\frac{x^{1-q}-1}{1-q}$, it is possible to generalize
the operation of multiplication introducing the $q$-product $\otimes_{q}$
such that $x\otimes_{q}y=exp_{q}\left[lq(x)+lq(y)\right]$  \cite{TSALLIS}
where \begin{equation}
exp_{q}(x)=\left[1+(1-q)x\right]^{\frac{1}{1-q}}.\label{eq:expqdef}\end{equation}

Taking into account that for the survival fraction we obtain expression
\eqref{eq:finalF} its analogy with \eqref{eq:expqdef} is evident.
Then, let us define the $exp_{\gamma}(x)$ function\begin{equation}
exp_{\gamma}(x)=\left[1+\frac{x}{\gamma}\right]^{\gamma}\label{eq:expgdef}\end{equation}
and the $l\gamma(x)$ its inverse function:\begin{equation}
l\gamma(exp_{\gamma}(x))=x.\label{eq:lgdef}\end{equation}

Let us introduce the $\gamma$-product of two numbers $x$ and $y$
as:\begin{equation}
x\otimes_{\gamma}y=exp_{\gamma}\left[l\gamma(x)+l\gamma(y)\right]=\left[x^{\frac{1}{\gamma}}+y^{\frac{1}{\gamma}}-1\right]^{\gamma}.\label{eq:gproddef}\end{equation}

Note that definitions \eqref{eq:expgdef} and \eqref{eq:lgdef} are
not essentially different from $e_{q}(x)$ and $lq(x)$. We are just
introducing these definitions to simplify the calculations.

\section{Results}

Let us now define the {}``effect potential'' as $\chi=-l\gamma(F_{s})$.
We demand this potential to satisfy the additive property. Then the
survival fraction expressed as\begin{equation}
F_{s}=exp_{\gamma}(-\chi)=exp_{\gamma}(-\frac{\gamma D}{\Delta}),\label{eq:fschi}\end{equation}
where the effect potential depends on $\left\langle E\right\rangle _{q}$,
becomes $\gamma$-multiplicative. This implies that the statistical
independence of the survival fractions is only possible when $\gamma\rightarrow\infty$
($q\rightarrow1$). 

Then the survival fraction for the sum of two potentials \emph{i.e.},
the survival fraction of a dose $B$ and a dose $A$ is:\begin{equation}
F_{s}(\chi_{A})\otimes_{\gamma}F_{s}(\chi_{B})=\left[1-\frac{\chi_{A}+\chi_{B}}{\gamma}\right]^{\gamma}\label{eq:fsprod}\end{equation}
and for $n$ doses\begin{equation}
F_{s}(n\chi)=\left[1-\sum_{i=1}^{n}\frac{\chi_{i}}{\gamma}\right]^{\gamma}=\left[1-\sum_{i=1}^{n}\frac{D_{i}}{\Delta_{i}}\right]^{\gamma}=exp_{\gamma}\left[-\sum_{i=1}^{n}\chi_{i}\right]=\left[\bigotimes_{i=1}^{n}\right]_{\gamma}F_{s}(\chi_{i})\label{eq:superprod}\end{equation}
where $\left[\bigotimes_{i=1}^{n}\right]_{\gamma}$ denotes the iterated
application of the $\gamma$-product.

\section{Discussion}

The obtained result permits to calculate directly the result of the
application of different radiation doses if $\gamma$ and $\Delta$
are known. Then the importance of experimental determination of $\gamma$
and $\Delta$ becomes evident. 

To determine $\Delta$ means to measure the minimal dose for which,
in a given tissue, no cell survives. This can be made by a progressive
application of a given kind of radiation. Once $\Delta$ is known
in those radiation conditions, a log-log plot of $F_{s}$ versus $1-D/\Delta$
must be fitted with the straight line whose slope is $\gamma$. Once
$\gamma$ is found the tissue is already characterized. Any other
kind of radiations or the same radiation applied in different conditions
to the same tissue can be fitted to the same straight line. 

This way, the introduction of the rules for generalization of the
product as in \cite{TSALLIS} permits an easy interpretation of the
properties of the survival fraction and a formula to calculate it
for several radiation doses: the survival fraction of several radiation
doses is the $\gamma$-product of the separate survival fractions
corresponding to each dose.

\section{Acknowledgements}

The authors wish to thank the Ministerio de Industria, Proyecto CD-TEAM,
CENIT. One of us (OSC) wants to acknowledge Prof. Ute Labonté and
Nike Fakiner for warm hospitality during his stay in Frankfurt where
this work was elaborated.

\bibliographystyle{unsrt}
\bibliography{short}

\end{document}